\documentclass[prl,twocolumn,superscriptaddress,preprintnumbers,amsmath,amssymb,floatfix,nofootinbib]{revtex4}

\usepackage{graphicx}
\usepackage{amsmath}    
\usepackage[normalem]{ulem}
\usepackage{bm}
\newcommand{\ket}[1]{\vert#1\rangle}
\newcommand{\bra}[1]{\langle#1\vert}
\def\opone{\leavevmode\hbox{\small1\kern-3.8pt\normalsize1}}

\usepackage[utf8x]{inputenc}
\usepackage{multirow} 
\usepackage{subfigure}
\usepackage{booktabs}
\usepackage{epsfig,graphics}
\usepackage{amsfonts,amssymb,amsmath}            
\usepackage{amsthm}

\newcommand*{\cA}{\mathcal{A}}

\newcommand{\eps}{\varepsilon}

\theoremstyle{plain}

\newtheorem{lemma}{Lemma}

\theoremstyle{definition}

\begin{document}
\title{An experimental test of all theories with predictive power beyond quantum theory}

\author{Terence E. \surname{Stuart}}
\email[]{testuart@ucalgary.ca} 
\affiliation{Institute for Quantum Information Science, and Department of Physics and Astronomy,  University of
  Calgary, 2500 University Drive NW, Calgary, Alberta T2N 1N4, Canada.}  

\author{Joshua A. \surname{Slater}}
\email[]{jslater@qis.ucalgary.ca} 
\affiliation{Institute for Quantum Information Science, and Department of Physics and Astronomy,  University of
  Calgary, 2500 University Drive NW, Calgary, Alberta T2N 1N4, Canada.}

\author{Roger \surname{Colbeck}}
\email[]{rcolbeck@perimeterinstitute.ca} 
\affiliation{Perimeter Institute for Theoretical Physics, 31 Caroline Street
   North, Waterloo, Ontario N2L 2Y5, Canada.}  

\author{Renato \surname{Renner}}
\email[]{renner@phys.ethz.ch} 
\affiliation{Institute for Theoretical Physics, ETH Zurich, 8093
Zurich, Switzerland.}

\author{Wolfgang \surname{Tittel}}
\email[]{wtittel@ucalgary.ca} 
\affiliation{Institute for Quantum Information Science, and Department of Physics and Astronomy,  University of
  Calgary, 2500 University Drive NW, Calgary, Alberta T2N 1N4, Canada.}

\begin{abstract}
According to quantum theory, the outcomes of future measurements
  cannot (in general) be predicted with certainty.  In some cases,
  even with a complete physical description of the system to be
  measured and the measurement apparatus, the outcomes of certain
  measurements are completely random.  This raises the question,
  originating in the paper by Einstein, Podolsky and Rosen \cite{EPR},
  of whether quantum mechanics is the optimal way to predict
  measurement outcomes.  Established arguments and experimental tests
  exclude a few specific alternative
  models \cite{Bell,GHZ1989,Leggett,FreedmanClauser,Aspect1982,Tittel1998,Weihs1998,Rowe2001,Aspect,Pan2000,GPKBZAZ,BLGKLS,Eisaman2008,BBGKLLS}. Here,
  we provide a complete answer to the above question, refuting any
  alternative theory with significantly more predictive power than
  quantum theory. More precisely, we perform various measurements on
  distant entangled photons, and, under the assumption that these
  measurements are chosen freely, we give an upper bound on how well
  any alternative theory could predict their outcomes \cite{CR_ext}.
  In particular, in the case where quantum mechanics predicts two
  equally likely outcomes, our results are incompatible with any
  theory in which the probability of a prediction is increased by more
  than $\sim$0.19. Hence, we can immediately refute any already
  considered or yet-to-be-proposed alternative model with more
  predictive power than this.
\end{abstract}

\maketitle

\begin{figure}
 \includegraphics[width=1\columnwidth]{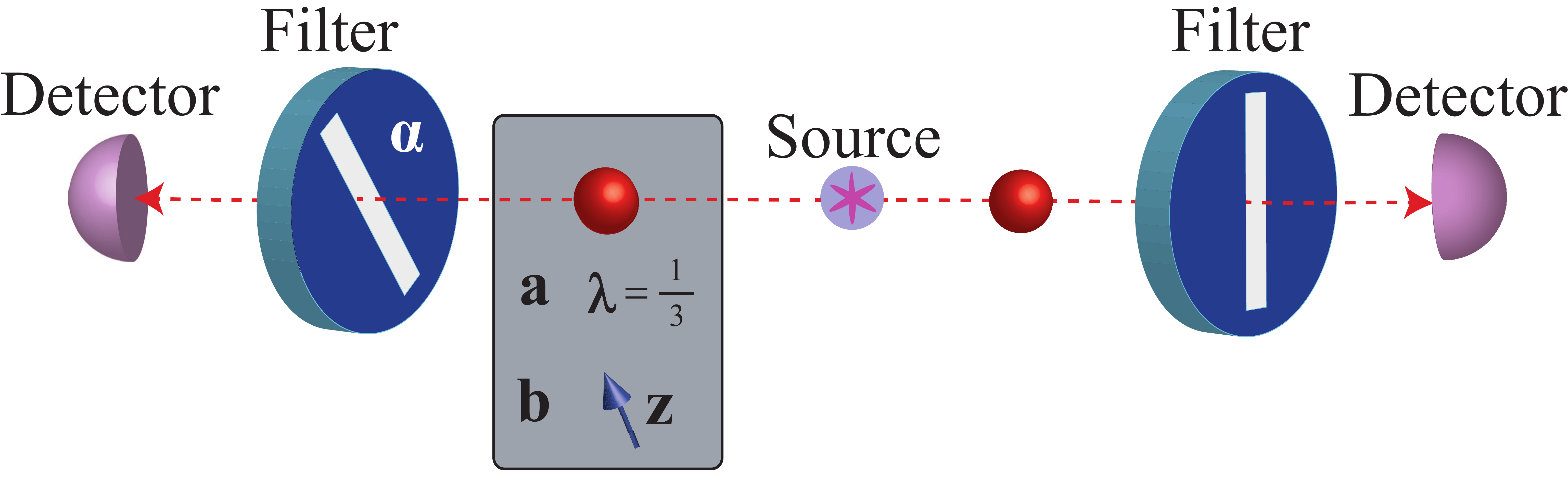}
   \caption{\label{fig:Leggett}{\bf Two alternative models.} Consider
     an experiment in which a source emits two spin-half particles
     travelling to two distant sites. Their spin direction is
     measured, e.g., by letting them pass through a filter that
     absorbs particles with the opposite spin direction. If the
     particles are initially maximally entangled, then the probability
     of correctly predicting whether the particle on the left is
     transmitted by the filter, which has a direction $\bm{\alpha}$, is,
     according to quantum mechanics, given by $p_{\mathrm{QM}} = 0.5$.
     {\bf a,} Bell's model of hidden variables \cite{Bell2} as an
     example for an alternative deterministic theory. Bell proposed a
     model in which the outcome of such a measurement is precisely determined by the particle's quantum
     mechanical state vector $\ket{\psi}$, the measurement specified by $\bm{\alpha}$, and an additional
     real number $\lambda$ -- a local hidden variable (LHV) that is not
     present in standard quantum mechanics. If we had access to
     $\lambda$, we could predict the outcome of the measurement with
     certainty, hence $p_{\mathrm{LHV}}=1$.  {\bf b,} A Leggett-type
     model \cite{Leggett} as an example for an alternative
     probabilistic theory.  Leggett imagined a theory in which each
     particle carries a hidden parameter, specified as a vector
     $\mathbf{{z}}$ that may be seen as a ``classical spin''. His
     model, adapted to general spin particles, prescribes that the
     probability that a particle with vector $\mathbf{{z}}$ is
     transmitted by a filter in direction $\bm{\alpha}$ is given by
     $\frac{1}{2} + \mathbf{{z}} \cdot \bm{\alpha}$. Since the vector
     $\mathbf{{z}}$ is unknown, we model it here as a random variable
     with no preferred direction (a detailed discussion of this and
     other distributions over $\mathbf{z}$ is deferred to the
      Appendix). A straightforward calculation then
     shows that, if we had access to the parameter $\mathbf{{z}}$, we
     could on average correctly predict the outcome with probability
     $p_{\mathrm{Leggett}}=0.75$.}
\end{figure}

Many of the predictions we make in everyday life are probabilistic.
Usually this is caused by having incomplete information, as is the
case when making weather forecasts.  On the other hand, even with all
the information available within quantum mechanics, the outcome of
certain experiments, e.g., the path taken by a spin-half particle in a
Stern-Gerlach experiment, is generally not predictable before the
start of the experiment.  This lack of predictive power has prompted a
long debate, which in turn led to important fundamental insights. In
particular, Kochen and Specker, and independently Bell, proved that
there cannot exist any \emph{noncontextual} theory that predicts
observations with certainty \cite{KS,Bell2}.  In a similar vein, Bell
showed \cite{Bell} that in general there cannot exist any additional
hidden property of the particle (a \emph{local hidden variable}) that
completely determines the outcome of any measurement on the
particle (for an illustration of such a model see Fig.~\ref{fig:Leggett}). Bell's argument relies on the fact that entangled particles
give rise to correlations that cannot be reproduced in any local
hidden variable theory. The existence of such correlations has been
confirmed in a series of increasingly sophisticated
experiments \cite{FreedmanClauser,Aspect1982,Tittel1998,Weihs1998,Rowe2001,Pan2000}.

The purpose of the above arguments was to refute theories in which
hidden parameters determine any experimental outcomes.  Access to
these parameters would allow us, in principle, to predict the outcomes
of any experiment with certainty. However, these arguments do not
preclude the possibility that we find a theory that has more
predictive power than quantum mechanics, while remaining
probabilistic. Consider again the Stern-Gerlach example where,
according to quantum mechanics, a particle entering the apparatus with
a certain spin orientation may be deviated in one of two directions,
each with probability 0.5. One may now conceive of a theory that,
depending on an additional parameter, would allow us to predict the
direction of deviation with a larger probability, say 0.75, thereby
improving the quantum mechanical prediction by 0.25. In
Fig.~\ref{fig:Leggett} we describe an example of such a theory, which
essentially corresponds to a proposal put forward by
Leggett \cite{Leggett}.

In this Letter we present experimental data that bounds the
probability, $\delta$, by which any alternative theory could improve
upon predictions made by quantum theory while still being consistent
with the assumption that measurement settings can be chosen freely. We
find that quantum theory is close to optimal in terms of its
predictive power.  Our work relies on a recent theoretical
argument \cite{CR_ext}, which is itself based on a sequence of
work \cite{pearle,BC,BHK,BKP,ColbeckRenner}, partly in the area of
quantum cryptography.  The experiment requires measuring bipartite
correlations of entangled particles, as well as establishing the
distributions of the associated individual measurement outcomes, for a
sufficiently large number of measurement settings. The maximum
increase of predictive power, $\delta$, of any alternative theory then
depends on the strength of the measured correlations, $I$, and on the
bias in the individual outcomes, $\nu$:
\begin{equation}
\delta\leq\frac{I}{2}+\nu.
\label{the_equation}
\end{equation}
\noindent
(These quantities are defined in the Appendix, where we also explain
the procedure for obtaining $I$ and $\nu$ from experimental
data.)

Before describing the experimental setup, let us briefly review the
main features of the theory leading to Eq.~\ref{the_equation} (a
complete derivation is given in the Appendix). Crucially, the framework used is operational, in the
sense that it refers only to directly observable quantities, such as
measurement outcomes.  For example, the Stern-Gerlach experiment
mentioned above outputs a binary value, $X$, indicating in
which direction the particle deviated. We associate with $X$ a time
coordinate $t$ and three spatial coordinates $(x_1, x_2, x_3)$,
corresponding to a point in spacetime where the value $X$ can be
observed. (If the value can be observed at different points in space
time, we may define further copies of $X$ with different spacetime
coordinates.) We note that these coordinates can be determined
operationally (e.g., using clocks and measuring rods, with respect to
a fixed reference system). We call such observable values with
spacetime coordinates \emph{spacetime variables (SVs)}.  In the same
manner, any parameter that is needed to specify the experiment (e.g.,
the orientation of the Stern-Gerlach apparatus) can be modelled as an
SV.

Consider now an experiment in which a spin measurement is made on a
particle that is maximally entangled with another one. According to
quantum theory, the outcome, $X$, of this measurement is random, even
with a complete description of the measurement apparatus,
$A$. However, an alternative theory may provide us with additional
information, $\Xi$ (which can also be modelled in terms of
SVs \cite{CR_ext}). We can then ask whether this additional information
$\Xi$ can be used to improve the predictions that quantum mechanics
makes about $X$, which depend on the measurement setting $A$ and the
initial state (which we assume to be fixed). This question has a
negative answer if the distribution of $X$, conditioned on $A$, is
unchanged when we learn $\Xi$. This can be expressed in terms of the
Markov chain condition \cite{CovThoMarkov},
\[
  X \leftrightarrow A \leftrightarrow \Xi 
\]
Equation~\ref{the_equation} now places a bound on the maximum
probability, $\delta$, by which this condition is violated. In other
words, the predictions obtained from quantum theory are optimal except
with probability (at most) $\delta$.

For the specific measurement described above, the validity of
Eq.~\ref{the_equation} relies only on the natural (and often implicit)
assumption that measurement parameters can be chosen freely. This
assumption can be expressed in the above framework as the requirement
that the SV corresponding to a measurement parameter, $A$, can be
chosen such that it is statistically independent of all SVs whose
coordinates lie outside the future lightcone of $A$ (Bell's theorem
also relies on such an assumption, for example, as explained in
Ref.~ \citenum{Bell3}).  When interpreted within the usual
relativistic spacetime structure, this is equivalent to demanding that
$A$ is uncorrelated with any pre-existing values in any frame.  We
also note that this requirement can be seen as a prerequisite for
non-contextuality, as pointed out in Ref.~ \citenum{ChenMontina} (where
an alternative proof that quantum theory cannot be extended, based on
the assumption of non-contextuality, is offered).

We note that our bound on the predictive power of alternative theories
can be extended to arbitrary measurements (not necessarily on
maximally entangled particles) if one makes one additional
assumption. This assumption is that the evolution of the state of a
physical system can always be correctly described by a unitary
operation if one includes part of the environment
in the description of the process \cite{CR_ext}.

\begin{figure*}
	\begin{center}
          \includegraphics[width=2\columnwidth]{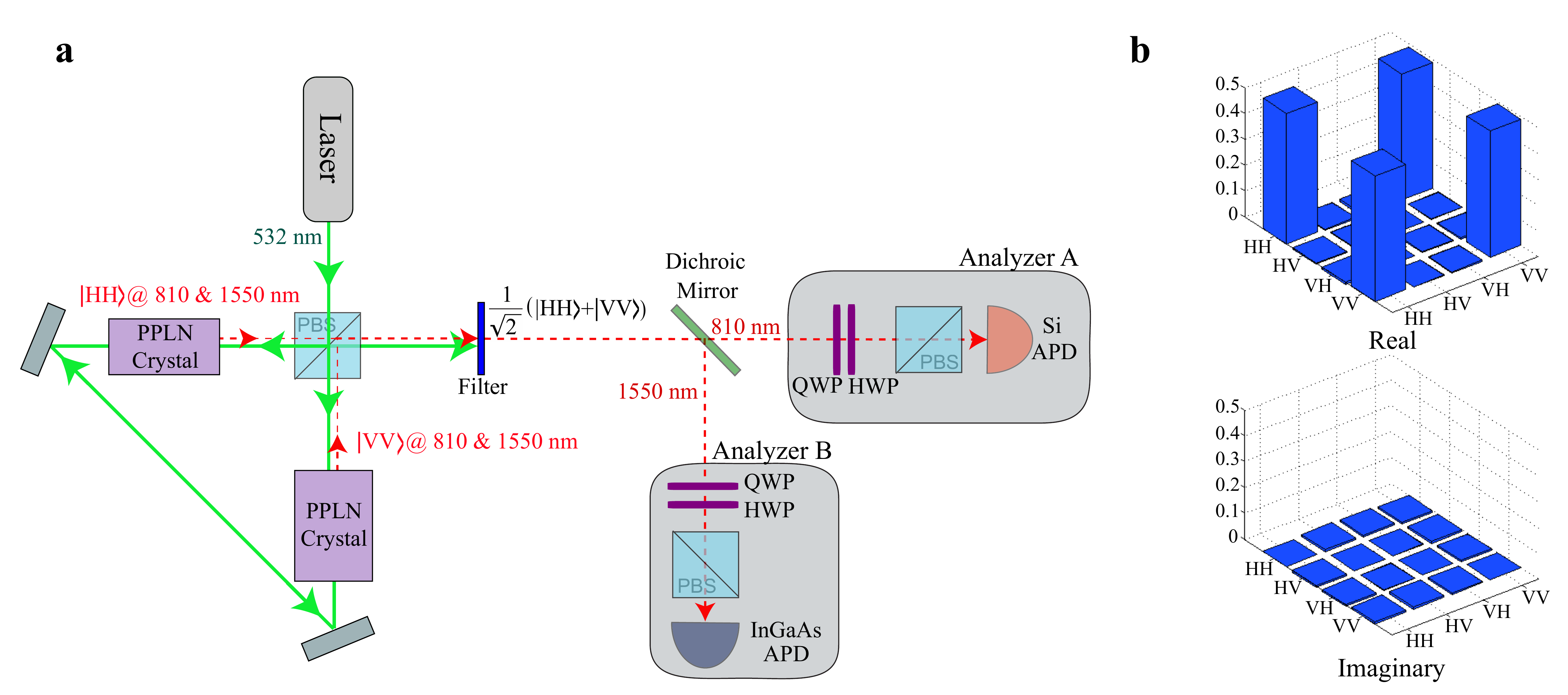}
          \caption{\label{fig:source}{\bf Generating and measuring
              entangled states.} \textbf{a}, Experimental setup. A
            diagonally polarized, continuous wave, 532~nm wavelength
            laser beam is split by a polarizing beam splitter (PBS)
            and travels both clockwise and counter-clockwise through a
            polarization Sagnac interferometer. The interferometer
            contains two type-I, periodically poled lithium niobate
            (PPLN) crystals configured to produce collinear,
            non-degenerate, 810/1550~nm wavelength photon pairs by
            means of spontaneous parametric down-conversion. As
            photon-pair generation is polarization dependent, the
            clockwise-travelling, vertically polarized
            (counter-clockwise travelling, horizontally polarized)
            pump light passes through the first crystal without
            interaction and may down-convert in the second crystal to
            produce two horizontally (vertically) polarized
            photons. For sufficiently small pump power, recombination
            of the two bi-photon modes on the PBS yields the entangled
            $\ket{\phi^+}$ state. After exiting the interferometer,
            the remaining pump light is filtered out using a high-pass
            filter. The entangled photons are separated on a dichroic
            mirror and sent to analyzers that allow one to measure
            polarization along any desired direction on the Bloch
            sphere. They consist of quarter wave plates (QWP), half
            wave plates (HWP), PBSs and single photon detectors. The
            810~nm photons are detected using a free-running Silicon
            avalanche photo-diode (Si APD), and 1550~nm photons are
            detected using an InGaAs APD triggered by detection events
            from the Si APD. \textbf{b}, Density matrix. Density
            matrix $\rho_{\mathrm{real}}$ of the bi-photon state
            produced by our source as calculated via
            maximum-likelihood quantum state
            tomography \cite{tomography} (see the Appendix for actual values). The fidelity,
            $F=\bra{\phi^+}\rho_{\mathrm{real}}\ket{\phi^+}$, between
            the detected state, $\rho_{\mathrm{real}}$, and the ideal
            state, $\ket{\phi^+}$, given by Eq.~\ref{eq:ourstate}, is
            (98.1$\pm$ 0.1)\%. }
	\end{center}
\end{figure*}

\begin{figure}
	\begin{center}
	\includegraphics[width=1\columnwidth]{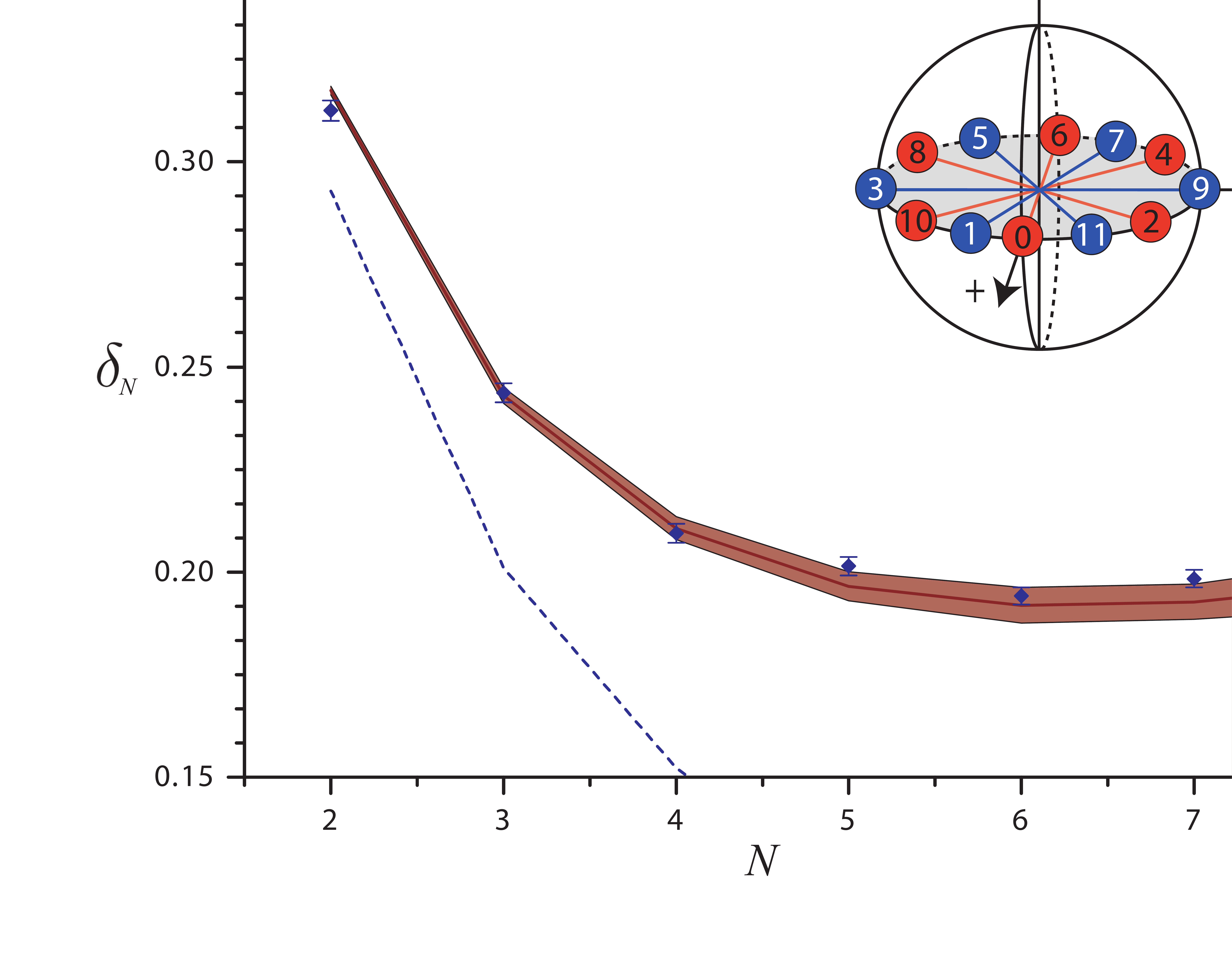}
          \caption{\label{fig:results}\textbf{Measurements and
            results. a,} Measurement settings. Graphical depiction of
          the polarization measurements along $\textbf{\textit{S}}_A$
          (red) and $\textbf{\textit{S}}_B$ (blue). The example shows
          $N=4$. \textbf{b,} Results.  Experimentally obtained values
          $\delta_N$ (blue diamonds) with one-standard-deviation
          uncertainties calculated from measurement results assuming
          Poissonian statistics.  Also shown is a curve joining the
          values predicted by quantum theory, including
          one-standard-deviation statistical uncertainties (solid red
          line and grey shaded area, respectively), calculated from
          the measured density matrix $\rho_{\mathrm{real}}$. The
          bounds of the shaded region are derived using Monte Carlo
          simulations and are consistent with the observed variations
          of the measured values. Finally, the dashed blue line is the
          theoretical curve, again calculated using quantum theory,
          that assumes the ideal $\ket{\phi^+}$ state, as in
          Eq.~\ref{eq:ourstate}, and perfect experimental apparatus
          with zero noise. It asymptotically approaches zero as $N$
          tends to infinity. For instance, for $N=6$ we find
          $\delta_6^{\mathrm{ideal}}=0.102$.
        }
	\end{center}
\end{figure}

The experimental setup we use to bound the quantities on the right-hand side
of Eq.~\ref{the_equation} is detailed in Fig.~\ref{fig:source}.  Our
source \cite{SSBT2011} generates photon pairs with high fidelity to the
entangled state
\begin{equation}
	\ket{\phi^+} = \dfrac{1}{\sqrt{2}}(\ket{HH} + \ket{VV}) \label{eq:ourstate},
\end{equation}
where $\ket{H}$ and $\ket{V}$ represent horizontal and vertical
polarization states, respectively, and replace the usual spin-up and
spin-down notation for spin-half particles. The photons from each pair
are separated and sent towards polarization analyzers that can be
adjusted to measure the polarization of an incoming photon along any
desired direction $\textbf{\textit{S}}=(S_+,S_L,S_H)$, where the
three-component vector $\textbf{\textit{S}}$ is expressed in terms of
its projections onto diagonal (+45$^\circ$), left-circular ($L$), and
horizontal ($H$) polarized components. $\textbf{\textit{S}}$ is conveniently
depicted on the Bloch sphere, see Fig \ref{fig:results}.

We perform a series of experiments that are parameterized by an
integer, $N$. Each experiment yields a value, $\delta_N=I_N/2+\nu_N$,
and we bound $\delta$ by the minimum over all measured $\delta_N$.
Each experiment comprises $N$ pairs of opposing spin measurement
settings per analyzer (i.e., $N$ different bases):
\begin{equation}
  \textbf{\textit{S}}_A(m) = \left(\begin{array}{ccc}\cos \left ({m\frac{\pi}{2N}}\right )  \\ \sin\left ({m\frac{\pi}{2N}}\right )\\0 \end{array}\right),  m \in \{0,  2, 4, \ldots , (4N-2)\}, 
 \end{equation}
\noindent
and 
\begin{equation}
  \textbf{\textit{S}}_B(n) = \left(\begin{array}{ccc}\cos \left ({n\frac{\pi}{2N}}\right )  \\ -\sin\left ({n\frac{\pi}{2N}}\right )\\0 \end{array}\right), n \in \{1,  3, 5, \ldots , (4N-1)\}.
 \end{equation}
\noindent
For each setting $\textbf{\textit{S}}_A(m)$, we count detected photons
over 80 sec to establish the bias $\nu_N$. Furthermore, for certain
joint measurements (described by specific combinations of
$\textbf{\textit{S}}_A(m)$ and $\textbf{\textit{S}}_B(n)$), we also
register the number of detected photon pairs over 40 sec to calculate
$I_N$, and hence $\delta_N$ (see the Appendix).

\begin{table}
\begin{center}
\begin{tabular}{ c || c | c || c}
  $N$ & $I_N$ & $\nu_N$ & $\delta_N$ \\
  \hline
  $2$ & $0.6196 \pm 0.0049$ & $0.0027 \pm 0.0003$ & $0.3125 \pm 0.0025$ \\
  $3$ & $0.4802 \pm 0.0046$ & $0.0036 \pm 0.0003$ & $0.2437 \pm 0.0023$ \\
  $4$ & $0.4103 \pm 0.0046$ & $0.0043 \pm 0.0003$ & $0.2094 \pm 0.0023$ \\
  $5$ & $0.3940 \pm 0.0045$ & $0.0045 \pm 0.0003$ & $0.2015 \pm 0.0023$ \\
  $6$ & $0.3791	\pm 0.0041$ & $0.0047 \pm 0.0003$ & $0.1942 \pm 0.0021$ \\
  $7$ & $0.3872 \pm 0.0042$ & $0.0048 \pm 0.0003$ & $0.1984 \pm 0.0021$ \\
\end{tabular}
\caption{\label{tab:results}{\bf Summary of Results.} The table
  shows values for $I_N$, bias $\nu_N$, as well as
  $\delta_N=I_N/2+\nu_N$. Statistical uncertainties (one standard
  deviation) are calculated from measurement results assuming
  Poissonian statistics.  }
 \end{center}
\end{table}

Our experimental results are depicted in Fig.~\ref{fig:results} and
summarized in Table~\ref{tab:results}.  We measured $\delta_N$ for
$N=2$ to $N=7$ and found the minimum, $\delta_6 = 0.194\pm 0.003$, for
$N=6$. Hence, the probability by which the predictions made by quantum
theory can be improved using any alternative theory is at most
$\sim$0.19.  As examples, note that this value rules out local hidden
variable as well as Leggett-type models (as explained in
Fig.~\ref{fig:Leggett} and the Appendix), since
$p_{\mathrm{LHV}} - p_{\mathrm{QM}} =0.5> \delta$ and
$p_{\mathrm{Leggett}} - p_{\mathrm{QM}} =0.25> \delta$, respectively.
(Here, $p$ denotes the maximum probability of correctly predicting the
measurement outcome in the model/theory indicated in the subscript.)  We remark that our experiments do not
close the locality and detection loopholes, so, strictly, the above
conclusions hold modulo the assumption that similar experiments
closing these loopholes would show the same results.

In conclusion, under the assumption that measurements can be chosen freely, no
theory can predict measurement outcomes substantially better than
quantum mechanics. In other words, any already considered or
yet-to-be-proposed theory that makes significantly better predictions
would either be incompatible with the experimental observations
presented herein, or be incompatible with our assumption that the
measurement parameters can be chosen freely. While the former is true,
for example, for local hidden variable theories (as already pointed
out by Bell \cite{Bell}) or for the Leggett model \cite{Leggett}, the de
Broglie-Bohm theory \cite{deBroglie,Bohm} is an example of the second
type|the theory cannot incorporate measurement parameters that satisfy
our free choice assumption.

\section{Acknowledgements}
  The authors thank F.~Bussi\`eres for help with setting up the
  photon pair source, and V.~Kiselyov for technical support. Research
  at Perimeter Institute is supported by the Government of Canada
  through Industry Canada and by the Province of Ontario through the
  Ministry of Research and Innovation. R.R. acknowledges support from
  the Swiss National Science Foundation (grant No.\ 200020-135048 and
  the NCCR QSIT) and from the European Research Council (grant No.\
  258932). W.T., T.E.S. and J.A.S. are supported by NSERC, QuantumWorks,
  General Dynamics Canada, iCORE (now part of Alberta Innovates), CFI,
  and AAET.

\newpage

\onecolumngrid

\section{Appendix}
	
\subsection{Calculation of bias, $\nu_N$, and correlation strength, $I_N$}

The bias is a measure of how close the distribution of the individual measurement outcomes, $X$, is to uniform. It is calculated from the number of 810 nm wavelength photons detected in opposing spin measurements.  In general, this number is not the same for all pairs of measurements. We take the bias, $\nu_N$, to be the maximum over the individual biases. Denoting the number of detected photons for setting $\textbf{\textit{S}}_A(m)$ by $M(m)$, we have:
\begin{equation}
  \nu_N = \dfrac{1}{2}\max_{a \in \{0,  2, 4, \ldots , (2N-2)\}} \left \{ \dfrac{ | M(a) - M(a+2N) |}{M(a) + M(a+2N)} \right \}.
 \end{equation}

The quantity $I_N$ measures the strength of the bipartite correlation between two analyzers. It is defined by
\begin{equation}
    I_N = P(0,2N-1) + \sum_{\genfrac{}{}{0pt}{}{a,b}{|a-b|=1}} \big(1-P(a,b) \big)
\end{equation}
where $a \in \{0, 2, 4, \ldots (2N-2)\}$, $b \in \{1, 3, 5, \ldots (2N-1)\}$. Furthermore, $P(a,b)$ is the sum of the probabilities for detecting two photons from a pair along the spin vectors $\textbf{\textit{S}}_A(a)$ and $\textbf{\textit{S}}_B(b)$, and along $-\textbf{\textit{S}}_A(a)$ and $-\textbf{\textit{S}}_B(b)$ (i.e. the probability of correlated outcomes):
\begin{equation}
    P(a,b) = \dfrac{M(a,b) + M(a+2N,b+2N)}{M(a,b)+M(a,b+2N)+M(a+2N,b)+M(a+2N,b+2N)} 
\end{equation}
where, e.g., $M(a,b)$ is the number of joint photon detections for measurements along $\textbf{\textit{S}}_A(a)$ and $\textbf{\textit{S}}_B(b)$.

\subsection{Proof of the bound}\label{app:1}
In this section, we prove the bound given in Equation~(1) in the main
text, which is stated as Lemma~\ref{lem:main} below.  We use a
bipartite scenario in which two spacelike separated measurements are
performed on a maximally entangled state.  We denote the choices of
observable $A\in\{0,2,\ldots,2N-2\}$ and $B\in\{1,3,\ldots,2N-1\}$ and
their outcomes $X\in\{+1,-1\}$ and $Y\in\{+1,-1\}$,
respectively\footnote{Note that the measurements we speak of in the
  Appendix have a slightly different form than those in the main
  text. Specifically, we now assume that measurements behave ideally,
  projecting onto one of two basis elements and leading to one of the two
  outcomes $\pm 1$.  In a real experiment, there is always the
  additional possibility of no photon detection (let us denote this
  outcome $0$).  The measurements discussed in the main text are
  configured to distinguish $+1$ from either $-1$ or $0$, or to
  distinguish $-1$ from either $+1$ or $0$. Both
  measurements are used in the experiment to infer the distribution of
  the ideal measurement with outcomes $\pm 1$.}.  We additionally
consider information that might be provided by an alternative theory
(this was denoted $\Xi$ in the main text), which is modelled as an
additional system with input $C$ and output $Z$~ \cite{CR_ext}.  If one
makes the assumption that the measurements can be chosen freely, then
the joint distribution $P_{XYZ|ABC}$ satisfies the non-signaling
conditions
\begin{align}
P_{XY|ABC}&=P_{XY|AB} \label{eq:NS1} \\
P_{XZ|ABC}&=P_{XZ|AC} \label{eq:NS2} \\
P_{YZ|ABC}&=P_{YZ|BC} \label{eq:NS3}
\end{align}
(see~ \cite{CR_ext} for a short proof of this).

Lemma~\ref{lem:main} gives a bound on the increase in predictive power
of any alternative theory in terms of the strength of correlations and
the bias of the individual outcomes.  The bound is expressed in terms
of the variational distance $D(P_Z,Q_Z) :=
\frac{1}{2}\sum_z|{P_Z(z)-Q_Z(z)}|$, which has the following
operational interpretation: if two distributions have variational
distance at most~$\delta$, then the probability that we ever notice a
difference between them is at most~$\delta$.

The bias is quantified by\footnote{A note on notation: we usually use
  lower case to denote particular instances of upper case random
  variables.}  $\nu_N:=\max_aD(P_{X|a},P_{\bar{X}})$, where
$P_{\bar{X}}$ is the uniform distribution on $X$.  To quantify the
correlation strength, we define
\begin{align} \label{eq:INdef}
I_N:=P(X=Y|A=0,B=2N-1)\ +
\sum_{\genfrac{}{}{0pt}{}{a,b}{|a-b|=1}} P(X \neq Y |A=a, B=b)  \ .
\end{align}
\noindent
This is equivalent to Equation~(6) in the main text.  We remark that
$I_N\geq 1$ is a Bell inequality, i.e.\ is satisfied by any local
hidden variable model.

\begin{lemma}\label{lem:main}
  For any non-signalling probability distribution, $P_{XYZ|ABC}$, we
  have
  \begin{align}\label{eq:markov}
    D(P_{Z|abcx},P_{Z|abc})&\leq \delta_N:=\frac{I_N}{2} + \nu_N
  \end{align}
  for all $a$, $b$, $c$, and $x$.
\end{lemma}
\noindent
To connect this back to the main text, we remark that the Markov chain
condition $X\leftrightarrow A\leftrightarrow \Xi$ is equivalent to
$P_{Z|abcx}=P_{Z|abc}$ (which corresponds to $\Xi$ not being of use to
predict $X$).  Hence, from the operational meaning of the variational
distance (given above), the left-hand-side of~\eqref{eq:markov}
corresponds to the maximum increase in the probability of correctly
predicting the outcome $X$, denoted $\delta$ in the main text.

The proof is an extension of an argument given in~ \cite{CR_ext} which
is based on \emph{chained Bell inequalities}~ \cite{pearle,BC} and
generalizes results of~ \cite{BHK,BKP,ColbeckRenner}.  Many steps of
this proof mirror those in~ \cite{CR_ext}, which we repeat for
completeness.  Furthermore, note that the bound derived in this Lemma
is tighter than that of~ \cite{CR_ext}.

\begin{proof}
  We first consider the quantity $I_N$ evaluated for the conditional
  distribution $P_{XY|AB,cz} =
  P_{XY|ABCZ}(\cdot,\cdot|\cdot,\cdot,c,z)$, for any fixed $c$ and
  $z$. The idea is to use this quantity to bound the variational
  distance between the conditional distribution $P_{X|a c z}$ and its
  negation, $1-P_{X|acz}$, which corresponds to the distribution of
  $X$ if its values are interchanged.  If this distance is small, it
  follows that the distribution $P_{X|acz}$ is roughly uniform.

  For $a_0:=0$, $b_0:=2N-1$, we have
\begin{align}
I_N(P_{XY|AB,cz})
&=
P(X=Y|A=a_0,B=b_0,C=c,Z=z)+\sum_{\genfrac{}{}{0pt}{}{a,b}{|a-b|=1}}P(X\neq Y|A=a,B=b,C=c,Z=z) \nonumber   \\
&\geq D(1-P_{X|a_0 b_0 c z},P_{Y|a_0 b_0 c z}) +
\sum_{\genfrac{}{}{0pt}{}{a,b}{|a-b|=1}}D(P_{X|abcz},P_{Y|abcz})
\nonumber\\
&= D(1-P_{X|a_0cz},P_{Y|b_0cz}) +
\sum_{\genfrac{}{}{0pt}{}{a,b}{|a-b|=1}}D(P_{X|acz},P_{Y|bcz})
\nonumber\\
&\geq D(1-P_{X|a_0cz},P_{X|a_0cz})\nonumber \\
&= 2D(P_{X|a_0b_0cz}, P_{\bar{X}}) \label{ItoD} \ .
\end{align}
The first inequality follows from the fact that $D(P_{X|\Omega},
P_{Y|\Omega}) \leq P(X \neq Y | \Omega)$ for any event $\Omega$ (a
short proof of this can be found in~ \cite{ColbeckRenner}).
Furthermore, we have used the non-signalling conditions
$P_{X|abcz}=P_{X|acz}$ (from~\eqref{eq:NS2}) and
$P_{Y|abcz}=P_{Y|bcz}$ (from~\eqref{eq:NS3}), and the triangle
inequality for $D$.  By symmetry, this relation holds for all $a$ and
$b$.  We hence obtain $D(P_{X|abcz},P_{\bar{X}})\leq\frac{1}{2}
I_N(P_{XY|AB,cz})$ for all $a$, $b$, $c$ and $z$.

We now take the average over $z$ on both sides of~\eqref{ItoD}. First, the
left hand side gives
\begin{align}
\sum_{z}P_{Z|abc}(z)I_N(P_{XY|AB,cz})&=\sum_{z}P_{Z|c}(z)I_N(P_{XY|AB,cz})\nonumber\\
&=\sum_{z}P_{Z|a_0b_0c}(z)P(X=Y|a_0, b_0, c,
z)+\sum_{\genfrac{}{}{0pt}{}{a,b}{|a-b|=1}}\sum_{z}P_{Z|abc}(z)P(X\neq
Y|a, b, c, z)\nonumber\\
&=P(X=Y|a_0, b_0, c)+\sum_{\genfrac{}{}{0pt}{}{a,b}{|a-b|=1}}P(X\neq
Y|a, b, c)\nonumber\\
&=I_N(P_{XY|AB,c}) \ ,
\end{align}
where we used the non-signalling condition $P_{Z|abc}=P_{Z|c}$ (which
is implied by~\eqref{eq:NS2} and~\eqref{eq:NS3}) several
times. Next, taking the average on the right hand side
of~\eqref{ItoD} yields
$\sum_zP_{Z|abc}(z)D(P_{X|abcz},P_{\bar{X}})=D(P_{XZ|abc},P_{\bar{X}}\times
P_{Z|abc})$, so we have
\begin{equation}\label{eq:8}
2D(P_{XZ|abc},P_{\bar{X}}\times
P_{Z|abc})\leq I_N(P_{XY|AB,c})=I_N(P_{XY|AB}).
\end{equation}
The last equality follows from the non-signalling
condition~\eqref{eq:NS1} (if $P(X=Y|a,b,c)$ or $P(X\neq Y|a,b,c)$
depended on $c$, then there would be signalling from $C$ to $A$ and
$B$).

Furthermore, note that
$$2D(P_{XZ|abc},P_{\bar{X}}\times P_{Z|abc})=\sum_{z}\bigl|P_{XZ|abc}(-1,z)-\frac{1}{2}P_{Z|abc}(z)\bigr|+\sum_{z}\bigl|P_{XZ|abc}(+1,z)-\frac{1}{2}P_{Z|abc}(z)\bigr|$$
and that both of the terms on the right hand side are equal (since
$P_{Z|abc}(z)=P_{XZ|abc}(-1,z)+P_{XZ|abc}(+1,z)$) i.e.\
$\sum_{z}\bigl|P_{XZ|abc}(x,z)-\frac{1}{2}P_{Z|abc}(z)\bigr|\leq\frac{I_N}{2}$
for all $a$, $b$, $c$ and $x$.  Note also that
$D(P_{X|a},P_{\bar{X}})=\bigl|P_{X|a}(x)-\frac{1}{2}\bigr|$ for all $x$.

Combining the above, we have
\begin{align*}
  D(P_{Z|abcx},P_{Z|abc})
& =
  \sum_z \bigl|\frac{1}{2} P_{Z|abcx}(z) - \frac{1}{2} P_{Z|abc}(z) \bigr| \\
& \leq
  \sum_z \bigl|\frac{1}{2} P_{Z|abcx}(z) - P_{X|abc}(x) P_{Z|abcx}(z) \bigr|
  + \sum_z \bigl| P_{X|abc}(x) P_{Z|abcx}(z) - \frac{1}{2} P_{Z|abc}(z) \bigr| \\
& =
  \sum_z P_{Z|abcx}(z) \bigl| \frac{1}{2} - P_{X|abc}(x) \bigr|
   + \sum_{z} \bigl| P_{XZ|abc}(x,z) - \frac{1}{2} P_{Z|abc}(z) \bigr| \\
& \leq D(P_{X|a},P_{\bar{X}})+\frac{I_N(P_{XY|AB})}{2}.
\end{align*}
This establishes the relation~\eqref{eq:markov}.
\end{proof}

\subsection*{Tightness}
We can also establish that this bound is tight, as follows.  Consider
a classical model in which, with probability $\eps$, we have $X=Y=Z=-1$,
and otherwise $X=Y=Z=+1$ (independently of $A$, $B$ and $C$).  This
distribution has $I_N(P_{XY|AB})=1$ and
$\nu=\frac{1}{2}-\eps$.  It also satisfies
$D(P_{Z|abcX=-1},P_{Z|abc})=1-\eps$, which is equal to the bound
implied by~\eqref{eq:markov}.

\subsection{Application to Leggett models}
In the Leggett model~ \cite{Leggett}, one imagines that improved
predictions about the outcomes for measurements on qubits are
available.  More precisely, each particle has an associated vector
(thought of as a hidden direction of its spin) and the outcome
distribution is expressed via the inner product with the vector
describing the measurement (see Figure~1 in the main text).  Denoting
the hidden vector for the first particle by ${\bf z}$, and its
measurement vector $\bm{\alpha}$ (this is the Bloch vector associated with
the chosen measurement direction), its outcomes are distributed
according to
\begin{equation}\label{eq:dotp}
P_{X|{\bm{\alpha}}{\bf {z}}}(\pm1)=\frac{1}{2}(1\pm{\bm{\alpha}}.{\bf z}).
\end{equation}
To relate this back to the discussion above, the Leggett model
corresponds to the case that there is no $C$, and where the hidden
vectors are contained in $Z$. Note that Leggett already showed his
model to be incompatible with quantum theory~ \cite{Leggett} and
experiments have since falsified it using specific
inequalities~ \cite{GPKBZAZ,BLGKLS,BBGKLLS}.  Here we discuss the model
in light of our experiment, which, it turns out, is sufficient to
falsify it.

Note that, as presented in~ \cite{Leggett} and above, the model is not
fully specified, since the distribution of the hidden vectors, ${\bf
  z}$, is not given.  In order to discuss the implications of our
experimental results, we refer to four cases (corresponding to
different distributions over ${\bf z}$).  Before describing these
cases, we first note that~\eqref{eq:markov} implies
\begin{equation}\label{eq:Zav}
  \langle
  D(P_{X|{\bf \bm{\alpha}z}},P_{\bar{X}})\rangle_{{\bf z}}\leq\delta_N,
\end{equation}
for all $\bm{\alpha}$, where $\langle\cdot\rangle_{{\bf z}}$ denotes the
expectation value over the vectors ${\bf z}$.  In order to
  falsify a particular version of the Leggett model, we compute
  $\delta_N^{\text{crit}}$, the smallest increase in predictive power
  under the assumption that a particular version of the Leggett model
  is correct (i.e.\ the smallest value of the left-hand-side
  of~\eqref{eq:Zav} over all $\bm{\alpha}$).  We then show that
  $\delta_N^{\text{crit}}$ is above the maximum increase in predictive
  power compatible with the experimental data, $\delta_N$, hence
  falsifying that version of Leggett's model.

{\bf First Case:} We imagine that the vector ${\bf z}$ is a fixed
vector (i.e.\ $P_{{\bf Z}}({\bf z})=1$) in the same plane on the Bloch
sphere as our measurements.  From~\eqref{eq:Zav} we find $D(P_{X|{\bf
    \bm{\alpha}z}},P_{\bar{X}})\leq\delta_N$ for all $\bm{\alpha}$. However,~\eqref{eq:dotp} implies $D(P_{X|{\bf
     \bm{\alpha}z}},P_{\bar{X}})=\frac{|{ \bm{\alpha}}.{\bf z}|}{2}$.  In order to make
$\max_{{\bm{\alpha}}}|{\bm{\alpha}}.{\bf z}|$ as small as possible, i.e.\
  find $\delta_N^{\text{crit1}}$, we require the vectors ${\bf z}$ to
be as far as possible from any of the possible $\bm{\alpha}$ vectors. If
the fixed vector ${\bf z}$ is in the plane containing the
measurements, this condition leads to $\max_{{ \bm{\alpha}}}|{ \bm{\alpha}}.{\bf
  z}|=\cos\frac{\pi}{2N}$ (i.e.\ ${\bf z}$ is positioned exactly in
between two settings for $\bm{\alpha}$).  Hence, this specific version of
the Leggett model is falsified if the measured
$\delta_N<\delta_N^{\text{crit1}}=\frac{1}{2}\cos\frac{\pi}{2N}$. As
shown in Table~\ref{tab:leggett1}, this is the case for all values of
$N$ assessed.

According to quantum theory, appropriately chosen measurements on a
maximally entangled state lead to correlations for which
$\delta_N=\frac{N}{2}(1-\cos\frac{\pi}{2N})$.  However, no
experimental realization can be noise-free, and this affects the
minimum $\delta_N$ attainable (see~ \cite{suarez,CR_ext}).  One way to
characterize the imperfection in the experiment is via the
visibility. In an experiment with visibility $V$, we instead obtain
$\delta_N=\frac{N}{2}(1-V\cos\frac{\pi}{2N})$, which for fixed $V$ has
a minimum at finite $N$.  In the case of this model, the minimum
visibility required to falsify it is 0.821 (with such a visibility the
model could be ruled out with $N=3$).

\begin{table}
\begin{center}
\begin{tabular}{ c || c | c |c| c|c|c}
  $N$ & $\delta_N^{\text{crit1}}$ &$\delta_N^{\text{crit2}}$ &$\delta_N^{\text{crit3}}$&$\delta_N^{\text{crit4}}$ & $\delta_N^{1}$ & $\delta_N^{2}$ \\
  \hline
 2 & {\bf 0.3536} &0.2& 0.25& 0.1768 &0.3125$\pm$0.0025\\
 3 & {\bf 0.4330} &{ 0.3062}&{\bf 0.25}&0.2165&0.2437$\pm$0.0023\\
 4 & {\bf 0.4619} &{ 0.3266}&{\bf 0.25}&{ 0.2310}&0.2094$\pm$0.0023\\
 5 & {\bf 0.4755} &{ 0.3362}&{\bf 0.25}&{ 0.2378}&0.2015$\pm$0.0023\\
 6 & {\bf 0.4830} &{\bf 0.3415}&{\bf 0.25}&{\bf 0.2415}&0.1942$\pm$0.0021& $ 0.2297 \pm 0.0020$\\
 7 &{\bf  0.4875} &{ 0.3447}&{\bf 0.25}&{ 0.2437}&0.1948$\pm$0.0021\\
\hline\hline
$V_{\min}$&0.821&0.906&0.946&0.951
\end{tabular}
\caption{\label{tab:leggett1}{\bf Leggett models: critical values
    and experimental data.} This table shows the critical values of
  $\delta_N$ required to rule out each of the four Leggett-type models
  discussed in the text.  Also shown are measured values for $\delta_N^1$ and $\delta_N^2$, where the superscript refers to measurements in the $\ket{+}-\ket{L}$ plane, and the $\ket{+}-\ket{H}$ plane of the Bloch sphere, respectively.  Bold values have
  $\delta_N^{1}<\delta_N^{\text{crit}}$ and, if required $\delta_N^{2}<\delta_N^{\text{crit}}$, i.e.\ they are ruled out
  by the data for that $N$. Note that the measurement in the second, orthogonal plane was only done for $N=6$.
  This value is relevant for ruling out the second and fourth cases.
  In the last row of the table, we note the minimum visibility
  required to rule out each of the four models.}
 \end{center}
\end{table}

{\bf Second case:} We now suppose ${\bf z}$ is a fixed vector, not
confined to the plane.  Then our basic measurements cannot strictly
\emph{rule out} this model: in principle, ${\bf z}$ could be close to
orthogonal to the plane containing the measurement vectors. (We remark
that if ${\bf z}$ is completely orthogonal to this plane, then it
would not be useful for making predictions.)  However, in order to
rectify this we can include a second set of measurements in the set of
random choices.  This set should be identical to the first apart from
being contained in an orthogonal plane.  We denote the sets $\cA_1$
and $\cA_2$ and we separately measure the $\delta_N$ values for each
plane, generating values denoted $\delta_N^1$ and $\delta_N^2$.
Analogously to the first case discussed above, this version of the
Leggett model is falsified unless for all ${\bm{\alpha}}\in\cA_1\cup\cA_2$,
$|{\bm{\alpha}}.{\bf z}|/2\leq\min(\delta_N^1,\delta_N^2)$.  In order to
make $\max_{{\bm{\alpha}}}|{\bm{\alpha}}.{\bf z}|$ as small as possible, we
require the vectors ${\bf z}$ to be as far as possible from any of the
possible $\bm{\alpha}$ vectors.  Consider now the four vectors
$(0,\sin\phi,\cos\phi)$, $(0,-\sin\phi,\cos\phi)$,
$(\cos\phi,\sin\phi,0)$ and $(\cos\phi,-\sin\phi,0)$ for
$\phi\leq\frac{\pi}{4}$ (these represent two neighbouring pairs of
measurement vectors (one in each plane), where we have chosen the
coordinates such that they are symmetric).  The vector equidistant
from these (in their convex hull) is
$(\frac{1}{\sqrt{2}},0,\frac{1}{\sqrt{2}})$.  It is then not possible
that for all ${\bm{\alpha}}\in\cA_1\cup\cA_2$, $|{\bm{\alpha}}.{\bf
  z}|/2\leq\min(\delta_N^1,\delta_N^2)$ provided
$\max(\delta_N^1,\delta_N^2)<\delta_N^{\text{crit2}}=\frac{1}{2\sqrt{2}}\cos\frac{\pi}{2N}$. As
shown in Table~\ref{tab:leggett1}, our experiment, which includes a measurement of $\delta_6$ in an orthogonal plane, also rules out this
version of the Leggett model.  (The minimum visibility required to
rule out this model is 0.906, which could do so using $N=4$.)

{\bf Third case:} We consider a slightly modified model in which ${\bf
  z}$ is distributed uniformly over the Bloch sphere.  This model is
arguably more natural since it is somewhat conspiratorial for ${\bf
  z}$ to always take a particular orientation with respect to the
measurements we perform (particularly if that measurement is chosen
freely), and is the one referred to in the main text.  In
this case, defining $\theta$ as the angle between $\bm{\alpha}$ and ${\bf
  z}$, we compute the left hand side of~\eqref{eq:Zav} as
$$\langle D(P_{X|{\bm{\alpha}z}},P_{\bar{X}})\rangle_{{\bf
    z}}=\int_{\theta=0}^{\pi}{\rm
  d}\theta\frac{|\cos\theta|\sin\theta}{4}=\frac{1}{4}.$$ 
This model is hence excluded if one finds
$\delta_N<\delta_N^{\text{crit3}}=\frac{1}{4}$ (measurements are needed only in
one plane). As shown in Table~\ref{tab:leggett1}, this is the case for
$N\geq 3$.  (The minimum visibility required to rule out this model is
0.946, which could do so for $N=5$.)

{\bf Fourth case:} Here we return to our measurements in two
orthogonal planes and ask whether our data is sufficient to falsify
the model for any distribution over ${\bf z}$.  (We can think of this
in terms of an adversarial picture. Suppose the set of possible
measurement choices is known to an adversary, who can pick the vector
${\bf z}$ according to any distribution he likes.  The aim is to show
that our measurement results are not consistent with any such
adversary.)  For this model to be correct we need
\begin{eqnarray*}
\frac{\langle|{\bm{\alpha}}.{\bf z}|\rangle_{{\bf
    z}}}{2}&\leq&\delta_N^1\text{ for all }{\bm{\alpha}}\in\cA_1\\
\frac{\langle|{\bm{\alpha}}.{\bf z}|\rangle_{{\bf
    z}}}{2}&\leq&\delta_N^2\text{ for all }{\bm{\alpha}}\in\cA_2.
\end{eqnarray*}
Again we can parameterize in terms of the four vectors introduced
previously.  When minimizing with respect to these four, we should
take $P_{{\bf Z}}$ to have support only on the set
$(\sin\theta,0,\cos\theta)$ (going off this line increases the inner
product with measurement vectors in both sets).  We thus have
\begin{equation*}
\langle|{\bm{\alpha}}.{\bf z}|\rangle_{{\bf z}}=\left\{\begin{array}{ll}\int_{\theta}{\rm
  d}\theta\rho(\theta)\cos\theta\cos\frac{\pi}{2N}&\text{ for all }{\bm{\alpha}}\in\cA_1\\
\int_{\theta}{\rm
  d}\theta\rho(\theta)\sin\theta\cos\frac{\pi}{2N}&\text{ for all }{\bm{\alpha}}\in\cA_2\end{array}\right.
\end{equation*}
where $\rho(\theta)$ is the probability density over $\theta$.

In other words, non-zero $\rho(\theta)$ gives contribution
$\cos\theta\cos\frac{\pi}{2N}$ to the first integral, and
$\sin\theta\cos\frac{\pi}{2N}$ to the second.  In order that both
integrals are equal, we should take $\rho(\theta)$ to be symmetric
about $\theta=\frac{\pi}{8}$.  For functions with this symmetry,
non-zero $\rho(\theta)$ gives contribution
$(\sin\theta+\cos\theta)\cos\frac{\pi}{2N}$ to both integrals.  The
minimum of this over $0\leq\theta\leq\frac{\pi}{8}$ is
$\cos\frac{\pi}{2N}$, which occurs for $\theta=0$.  It follows that
the most experimentally challenging distribution to rule out is
$\rho(\theta)=\frac{1}{2}(\delta_{\theta,0}+\delta_{\theta,\frac{\pi}{4}})$,
where $\delta_{x,y}$ is the Kronecker delta (this being the
distribution that requires the lowest measured $\delta_N$ to
eliminate).  For this distribution, we have $\max_{{\bm{\alpha}}}\langle|{\bm{\alpha}}.{\bf z}|\rangle_{{\bf
    z}}/2=\frac{1}{4}\cos\frac{\pi}{2N}$, so this model is ruled out
for
$\max(\delta_N^1,\delta_N^2)<\delta_N^{\text{crit4}}=\frac{1}{4}\cos\frac{\pi}{2N}$.
Again, as detailed in Table~\ref{tab:leggett1}, our experimental data
is sufficient to do so. (The lowest visibility that could rule out
this case is $0.951$, which would do so for $N=5$).

\subsection*{Comment on minimum visibilities required to rule out
  Leggett models
}
Here we briefly compare the visibilities required to rule out Leggett
models using our approach with those needed in previously considered
Leggett inequalities.  We remind the reader that the technique used in
the present work generates conclusions that apply to arbitrary
theories and were not developed with Leggett's model in mind.
Nevertheless, use of this new approach to rule out Leggett models
requires comparable visibilities to those of previously discussed
inequalities.  More specifically, the claimed minimum visibilities are
0.974 in Gr\"oblacher \textit{et al.}~ \cite{GPKBZAZ} and 0.943 for the
alternative inequality of Branciard {\it et
  al.}~ \cite{BLGKLS,BBGKLLS}, which is only slightly below the value
we require to rule out all of the four models above.

We note that the visibility for measurements in the plane used in the
main text was 0.967 $\pm$ 0.007, while the visibility in the
orthogonal plane (measured for the purposes of ruling out the second
and fourth cases) was 0.977 $\pm$ 0.009.

\begin{figure}[h]
\includegraphics[width=0.32\textwidth]{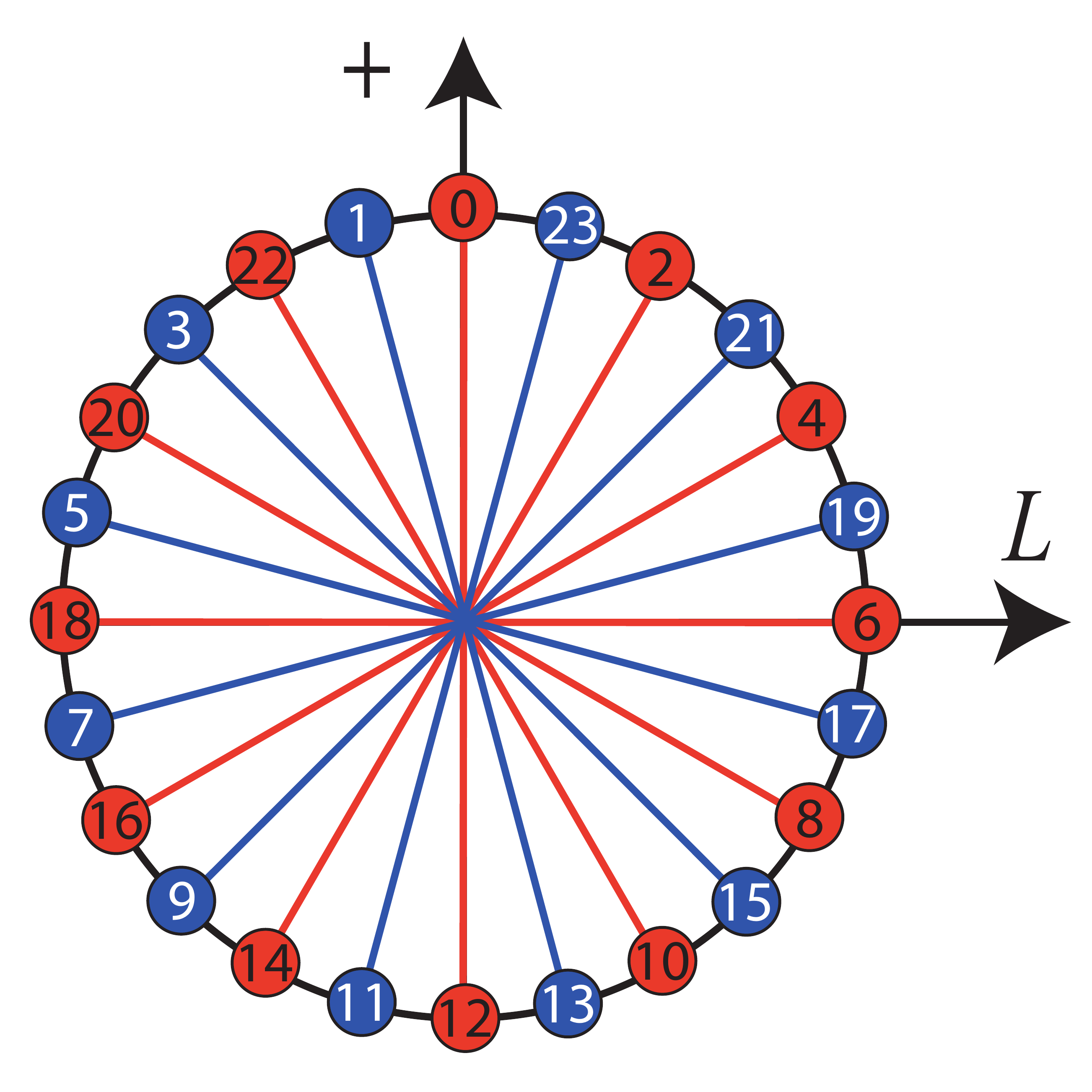}
\caption{\label{fig:settings_Neq6}{\bf Measurement settings for
    $N=6$.}  All settings are in the $\ket{+}$-$\ket{L}$ plane in the
  Bloch sphere.  Alice's settings are indicated in red and Bob's in
  blue.}
\end{figure}

\subsection{Density Matrix and Raw Data}

The experimental settings as well as the associated measurement
results that allow reconstruction of the density matrix are given in
Table~\ref{tab:tomodata}.  The most likely density matrix is detailed
in Table~\ref{tab:matrix}.  Note that this density matrix is not used
for the calculation of experimental values for $\delta_N$, $I_N$ or
$\nu_N$, but is included to characterize our source.
The measurements settings used to experimentally determine $\delta_6$ are depicted in
Figure~\ref{fig:settings_Neq6}, and Table~\ref{tab:raw_delta} lists the
results used to calculate $\delta_6$ from the bi-partite
correlation $I_6$ and the bias $\nu_6$.

\begin{table}[h]
  \begin{center}
  \footnotesize 
  \begin{tabular}{ c  c | c c c c |  r   r   } 
  \toprule
  \multicolumn{2}{c}{Setting} & \multicolumn{1}{|c}{$HWP_A$}  & \multicolumn{1}{c}{$QWP_A$}  & \multicolumn{1}{c}{$HWP_B$}  & \multicolumn{1}{c|}{$QWP_B$} & \multicolumn{1}{c}{$R_C$} & \multicolumn{1}{c}{$\Delta R_C$}\\
  \multicolumn{1}{c}{$a$} & \multicolumn{1}{c}{$b$}  & \multicolumn{1}{|c}{(\textdegree)}  & \multicolumn{1}{c}{(\textdegree)}  & \multicolumn{1}{c}{(\textdegree)} & \multicolumn{1}{c|}{(\textdegree)} & \multicolumn{1}{c}{(cps)} & \multicolumn{1}{c}{(cps)}  \\ \hline
$\ket{H}$ & $\ket{H}$  & 	0 & 	0 & 	0 & 	0 & 	240.0 & 2.8  \\
$\ket{H}$ &  $\ket{V}$  & 	0& 	0& 	45&	0 & 	1.8 & 0.2 \\
$\ket{H}$ &  $\ket{+}$ &	0 & 	0&  22.5&  	45& 	118.4 & 2.0  \\
$\ket{H}$ &  $\ket{- }$ & 	0 & 	0& -22.5& 	45&	125.3 & 2.0  \\
$\ket{H}$ &  $\ket{R}$ &	0 &	0& 	0& 	45& 	118.7 & 2.0  \\
$\ket{H}$ &  $\ket{L}$ &	0 & 	0& 	0& 	-45& 	130.6 & 2.0 \\
$\ket{V}$ &  $\ket{H}$ & 	45& 	0& 	0& 	0& 	2.0 & 0.3  \\
$\ket{V}$ &  $\ket{V}$ & 	45& 	0& 	45& 	0& 	230.6 & 2.8  \\
$\ket{V}$ &  $\ket{+}$ & 	45& 	0& 	22.5& 45& 	119.1 & 2.0 \\
$\ket{V}$ &  $\ket{- }$& 	45& 	0& 	-22.5& 45& 	118.5 & 2.0 \\
$\ket{V}$ &  $\ket{R}$ & 	45& 	0& 	0& 	 45& 	123.8 & 2.0 \\
$\ket{V}$ &  $\ket{L }$&  	45& 	0& 	0& 	-45& 	112.4 & 1.9 \\
$\ket{+}$ &  $\ket{H}$ & 	22.5&	45& 	0& 	0& 	115.7 & 2.0  \\
$\ket{+}$ &  $\ket{V}$ & 	22.5&	45& 	45& 	0& 	122.0 & 2.0  \\
$\ket{+}$ &  $\ket{+}$ & 	22.5&	45& 	22.5&	 45& 	245.8 & 2.9 \\
$\ket{+}$ &  $\ket{-}$ & 	22.5&	45& 	-22.5& 45& 	3.6 & 0.3  \\
$\ket{+}$ &  $\ket{R}$ & 	22.5&	45& 	0& 	 45& 	118.8 & 2.0  \\
$\ket{+}$ &  $\ket{L}$ & 	22.5&	45& 	0& 	-45& 	111.3 & 1.9 \\
$\ket{- }$&  $\ket{H }$&  	-22.5&45& 	0& 	0& 	124.2 & 2.0  \\
$\ket{- }$&  $\ket{V }$&  	-22.5&45& 	45& 	0& 	116.6 & 2.0 \\
$\ket{- }$&  $\ket{+}$ & 	-22.5&45& 	22.5&  45& 	3.7 & 0.4 \\
$\ket{- }$&  $\ket{- }$&  	-22.5&45& 	-22.5& 45& 	241.1 & 2.8  \\
$\ket{- }$&  $\ket{R }$&  	-22.5&45& 	0& 	 45& 	124.5 & 2.0  \\
$\ket{- }$&  $\ket{L }$&  	-22.5&45& 	0& 	-45& 	138.5 & 2.1 \\
$\ket{R}$ & $\ket{H }$&  	0& 	45& 	0& 	0& 	112.7 & 1.9  \\
$\ket{R}$ &  $\ket{V}$ & 	0& 	45& 	45& 	0& 	124.5 & 2.0 \\
$\ket{R}$ &  $\ket{+}$ & 	0& 	45& 	22.5&  45& 	119.2 & 2.0  \\
$\ket{R}$ &  $\ket{- }$& 	0& 	45& 	-22.5& 45& 	121.7 & 2.0 \\
$\ket{R}$ &  $\ket{R}$ & 	0& 	45& 	0& 	 45& 	2.9 & 0.3 \\
$\ket{R}$ &  $\ket{L}$ & 	0& 	45& 	0& 	-45& 	239.3 & 2.8  \\
$\ket{L }$&  $\ket{H}$ & 	0& 	-45& 	0& 	0& 	128.8 & 2.1  \\
$\ket{L }$&  $\ket{V }$& 	0& 	-45& 	45& 	0& 	109.0 & 1.9  \\
$\ket{L }$&  $\ket{+}$ &  	0& 	-45& 	22.5&  45& 	115.8 & 2.0  \\
$\ket{L }$&  $\ket{- }$&  	0& 	-45& 	-22.5& 45& 	124.6 & 2.0  \\
$\ket{L }$&  $\ket{R}$ & 	0& 	-45& 	0& 	 45& 	230.8 & 2.8  \\
$\ket{L }$&  $\ket{L }$&  	0& 	-45& 	0& 	-45& 	4.0 & 0.4 \\ \hline
    \bottomrule
  \end{tabular}
  \normalsize
  \caption{\label{tab:tomodata}{\bf Tomographic Data.} This table
    shows raw data collected to find the density matrix shown in
    Table~\ref{tab:density}.  The coincidence rates between the Si avalanche photodiode (APD)
    and the triggered 1550 nm InGaAs APD ($R_C$) for each set of
    photon analyzer settings are given in average counts per second
    (cps), as are their one standard deviation uncertainties ($\Delta
    R_C$).  Settings $a$ and $b$ were implemented using one quarter
    wave plate followed by one half wave plate in each analyzer.
    These waveplates were set at angles $HWP_A$, $QWP_A$, $HWP_A$, and
    $QWP_A$.  Data collection time for each point was 30 seconds. }
  \end{center}
\end{table}

\begin{table}[!ht]
\centering
\subtable[\ $\rho_{Re}$]{
\footnotesize
\centering
\begin{tabular}{r | c  c  c  c  } 
\multicolumn{5}{c}{} \\
 \multicolumn{1}{r}{} &  \multicolumn{1}{c}{$\bra{HH}$} & \multicolumn{1}{c}{$\bra{HV}$} & \multicolumn{1}{c}{$\bra{VH}$} & \multicolumn{1}{c}{$\bra{VV}$} \\ \cline{2-5}
 $\ket{HH}$ & 0.5038 & -0.0052 & -0.0092 & 0.4851 \\ 
 $\ket{HV}$ & -0.0052 & 0.0040 & 0.0001 & -0.0011 \\ 
 $\ket{VH}$ & -0.0092 & 0.0001 & 0.0043 & -0.0044 \\ 
 $\ket{VV}$ & 0.4851 & -0.0011 & -0.0044 & 0.4879 \\ 
\end{tabular}
\label{tab:chapter4:1a}
}
\subtable[\ $\rho_{Im}$]{
\footnotesize
\centering
\begin{tabular}{r | c  c  c  c  } 
\multicolumn{5}{c}{} \\
 \multicolumn{1}{r}{} &  \multicolumn{1}{c}{$\bra{HH}$} & \multicolumn{1}{c}{$\bra{HV}$} & \multicolumn{1}{c}{$\bra{VH}$} & \multicolumn{1}{c}{$\bra{VV}$} \\ \cline{2-5}
 $\ket{HH}$ & 0.0000 & 0.0155 & 0.0138 & -0.0140 \\ 
 $\ket{HV}$ & -0.0155 & 0.0000 & 0.0017 & -0.0156 \\ 
 $\ket{VH}$ & -0.0138 & -0.0017 & 0.0000 & -0.0113 \\ 
 $\ket{VV}$ & 0.0140 & 0.0156 & 0.0113 & 0.0000 \\ 
\end{tabular}
\label{tab:chapter4:1b}
}
\caption{\label{tab:density}{\bf Density matrix.}  The real and imaginary
  parts of the density matrix generated by maximum likelihood quantum state tomography.}
\label{tab:matrix}
\end{table}

\begin{table}[h]
  \begin{center}
  \footnotesize 
  \begin{tabular}{ c  c | c   c | r  r | r  r  r | r  r  } 
  \toprule
  \multicolumn{2}{c}{Setting} & \multicolumn{1}{|c}{$HWP_A$} & \multicolumn{1}{c}{$HWP_B$} & \multicolumn{1}{|c}{$R_{Si}$} & \multicolumn{1}{c}{$R_{C}$} & \multicolumn{1}{|c}{$1-P{(m,n)}$} & \multicolumn{1}{c}{$P{(m,n)}$} & \multicolumn{1}{c}{$\Delta P(m,n)$} & \multicolumn{1}{|c}{$\nu$} & \multicolumn{1}{c}{$\Delta \nu$} \\
  \multicolumn{1}{c}{$m$} & \multicolumn{1}{c}{$n$} & \multicolumn{1}{|c}{(\textdegree)} & \multicolumn{1}{c}{(\textdegree)} & \multicolumn{1}{|c}{(cps)} & \multicolumn{1}{c}{(cps)} & \multicolumn{1}{|c}{} & \multicolumn{1}{c}{} & \multicolumn{1}{c}{} & \multicolumn{1}{|c}{} & \multicolumn{1}{c}{} \\ \hline
    0 & 11 & 0 & -41.25 & 18692 & 6.7 &  &  &  &  &    \\
    0 & 23 & 0 & -86.25 & 18597 & 264.9 &  &  &  &  &    \\
    12 & 23 & -45 & -86.25 & 19017 & 7.5 &  &  &  &  &   \\
    12 & 11 & -45 & -41.25 & 18976 & 270.6 & \multirow{-4}{*}{0.0259} & \multirow{-4}{*}{0.9741} & \multirow{-4}{*}{0.0011} & \multirow{-4}{*}{0.0047} & \multirow{-4}{*}{0.0003}  \\ \hline
    0 & 1 & 0 & -3.75 & 18552 & 267.3 &  &  &  &  &   \\
    0 & 13 & 0 & -48.75 & 18588 & 9.2 &  &  &  &  &   \\
    12 & 13 & -45 & -48.75 & 18919 & 263.0 &  &  &  &  &   \\
    12 & 1 & -45 & -3.75 & 18900 & 9.6 & \multirow{-4}{*}{0.9657} & \multirow{-4}{*}{0.0343} & \multirow{-4}{*}{0.0012} & \multirow{-4}{*}{0.0045} & \multirow{-4}{*}{0.0003} \\ \hline
    2 & 1 & -7.5 & -3.75 & 18571 & 267.8 &  &  &  &  &    \\
    2 & 13 & -7.5 & -48.75 & 18632 & 7.6 &  &  &  &  &    \\
    14 & 13 & -52.5 & -48.75 & 18772 & 263.3 &  &  &  &  &    \\
    14 & 1 & -52.5 & -3.75 & 19018 & 9.0 & \multirow{-4}{*}{0.9697} & \multirow{-4}{*}{0.0303} & \multirow{-4}{*}{0.0012} & \multirow{-4}{*}{0.0039} & \multirow{-4}{*}{0.0003}  \\ \hline
    2 & 3 & -7.5 & -11.25 & 18528 & 266.8 &  &  &  &  &    \\
    2 & 15 & -7.5 & -56.25 & 18746 & 8.5 &  &  &  &  &    \\
    14 & 15 & -52.5 & -56.25 & 18910 & 268.4 &  &  &  &  &   \\
    14 & 3 & -52.5 & -11.25 & 18990 & 10.1 & \multirow{-4}{*}{0.9665} & \multirow{-4}{*}{0.0335} & \multirow{-4}{*}{0.0012} & \multirow{-4}{*}{0.0042} & \multirow{-4}{*}{0.0003}   \\ \hline
    4 & 3 & -15 & -11.25 & 18712 & 271.2 &  &  &  &  &    \\
    4 & 15 & -15 & -56.25 & 18604 & 7.9 &  &  &  &  &    \\
    16 & 15 & -60 & -56.25 & 18430 & 263.6 &  &  &  &  &   \\
    16 & 3 & -60 & -11.25 & 18449 & 7.6 & \multirow{-4}{*}{0.9720} & \multirow{-4}{*}{0.0280} & \multirow{-4}{*}{0.0011} & \multirow{-4}{*}{0.0029} & \multirow{-4}{*}{0.0003}  \\ \hline
    4 & 5 & -15 & -18.75 & 18058 & 262.4 &  &  &  &  &    \\
    4 & 17 & -15 & -63.75 & 17979 & 7.8 &  &  &  &  &    \\
    16 & 17 & -60 & -63.75 & 18147 & 254.3 &  &  &  &  &    \\
    16 & 5 & -60 & -18.75 & 18201 & 9.8 & \multirow{-4}{*}{0.9671} & \multirow{-4}{*}{0.0329} & \multirow{-4}{*}{0.0012} & \multirow{-4}{*}{0.0022} & \multirow{-4}{*}{0.0003}  \\ \hline
    6 & 5 & -22.5 & -18.75 & 18034 & 262.0 &  &  &  &  &   \\
    6 & 17 & -22.5 & -63.75 & 18129 & 7.8 &  &  &  &  &    \\
    18 & 17 & -67.5 & -63.75 & 18045 & 261.9 &  &  &  &  &   \\
    18 & 5 & -67.5 & -18.75 & 18166 & 7.5 & \multirow{-4}{*}{0.9716} & \multirow{-4}{*}{0.0284} & \multirow{-4}{*}{0.0011} & \multirow{-4}{*}{0.0003} & \multirow{-4}{*}{0.0003}   \\ \hline
    6 & 7 & -22.5 & -26.25 & 18044 & 259.2 &  &  &  &  &    \\
    6 & 19 & -22.5 & -71.25 & 18499 & 8.5 &  &  &  &  &    \\
    18 & 19 & -67.5 & -71.25 & 18438 & 257.9 &  &  &  &  &   \\
    18 & 7 & -67.5 & -26.25 & 18360 & 11.1 & \multirow{-4}{*}{0.9634} & \multirow{-4}{*}{0.0366} & \multirow{-4}{*}{0.0013} & \multirow{-4}{*}{0.0017} & \multirow{-4}{*}{0.0003}   \\ \hline
    8 & 7 & -30 & -26.25 & 18354 & 261.9 &  &  &  &  &   \\
    8 & 19 & -30 & -71.25 & 18386 & 7.3 &  &  &  &  &   \\
    20 & 19 & -75 & -71.25 & 18317 & 262.8 &  &  &  &  &    \\
    20 & 7 & -75 & -26.25 & 18403 & 7.0 & \multirow{-4}{*}{0.9735} & \multirow{-4}{*}{0.0265} & \multirow{-4}{*}{0.0011} & \multirow{-4}{*}{0.0002} & \multirow{-4}{*}{0.0003}   \\ \hline
    8 & 9 & -30 & -33.75 & 18305 & 261.7 &  &  &  &  &    \\
    8 & 21 & -30 & -78.75 & 18254 & 8.5 &  &  &  &  &   \\
    20 & 21 & -75 & -78.75 & 18066 & 256.7 &  &  &  &  &   \\
    20 & 9 & -75 & -33.75 & 18200 & 9.8 & \multirow{-4}{*}{0.9659} & \multirow{-4}{*}{0.0341} & \multirow{-4}{*}{0.0012} & \multirow{-4}{*}{0.0020} & \multirow{-4}{*}{0.0003}  \\ \hline
    10 & 9 & -37.5 & -33.75 & 18042 & 260.4 &  &  &  &  &    \\
    10 & 21 & -37.5 & -78.75 & 18102 & 7.3 &  &  &  &  &    \\
    22 & 21 & -82.5 & -78.75 & 18100 & 257.1 &  &  &  &  &   \\
    22 & 9 & -82.5 & -33.75 & 18073 & 9.0 & \multirow{-4}{*}{0.9696} & \multirow{-4}{*}{0.0304} & \multirow{-4}{*}{0.0012} & \multirow{-4}{*}{0.0002} & \multirow{-4}{*}{0.0003}   \\ \hline
    10 & 11 & -37.5 & -41.25 & 17979 & 256.3 &  &  &  &  &    \\
    10 & 23 & -37.5 & -86.25 & 17958 & 9.6 &  &  &  &  &   \\
    22 & 23 & -82.5 & -86.25 & 17857 & 256.3 &  &  &  &  &   \\
    22 & 11 & -82.5 & -41.25 & 18014 & 10.9 & \multirow{-4}{*}{0.9615} & \multirow{-4}{*}{0.0385} & \multirow{-4}{*}{0.0013} & \multirow{-4}{*}{0.0005} & \multirow{-4}{*}{0.0003}   \\ \hline
    \bottomrule
  \end{tabular}
  \normalsize
  \caption{\label{tab:raw_delta}{\bf Raw Data used to calculate
      $\delta_6^1$.} This table shows raw data collected to find
    $\delta_6^1=0.1942\pm 0.0021$.  $HWP_A$ and $HWP_B$ are the half
    wave-plate settings that, together with quarter waveplate settings
    of $-45^{\circ}$ on side A and $+45^{\circ}$ on side B, realize
    the measurements corresponding to $m$ and $n$ as shown in
    Figure~\ref{fig:settings_Neq6}.  The free running Silicon APD
    rates ($R_{Si}$) and the coincidence rates between the Si APD and
    the triggered 1550 nm InGaAs APD ($R_C$) are both given in
    average counts per second.  $P(m,n)$ is the probability of
    correlated outcomes and $\nu$ is the bias for individual
    measurements as detailed above.  Data collection
    time for each point was 40 seconds.  Uncertainties are one standard
    deviation.}
  \end{center}
\end{table}

\end{document}